\documentclass{article}

\usepackage{arxiv}

\usepackage[utf8]{inputenc} 
\usepackage[T1]{fontenc}    
\usepackage{hyperref}       
\usepackage{url}            
\usepackage{booktabs}       
\usepackage{amsfonts}       
\usepackage{nicefrac}       
\usepackage{microtype}      
\usepackage{lipsum}		
\usepackage{graphicx}
\usepackage{natbib}
\usepackage{doi}

\title{The Power of 10: New Rules for the Digital World}

\date{Submitted 26 June 2025, revised 29 August 2025, accepted 3 November 2025}	

\author{
  Sarah Spiekermann-Hoff\thanks{Corresponding author} \\
  Vienna University of Economics and Business\\
  Vienna, Austria \\
  \And
  \href{https://orcid.org/0000-0002-8834-7388}{\includegraphics[scale=0.06]{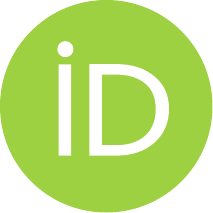}\hspace{1mm}Marc Langheinrich} \\
  Universit\`a della Svizzera italiana\\
  Lugano, Switzerland \\
  \And
  Johannes Hoff \\
  University of Innsbruck\\
  Innsbruck, Austria \\
  \And
  Christiane Wendehorst \\
  University of Vienna\\
  Vienna, Austria \\
  \And
  J\"urgen Pfeffer \\
  Technical University Munich\\
  Munich, Germany \\
  \And
  Thomas Fuchs \\
  Heidelberg University Hospital\\
  Heidelberg, Germany \\
  \And
  Armin Grunwald \\
  Karlsruhe Institute of Technology\\
  Karlsruhe, Germany \\
}

\renewcommand{\headeright}{Preprint}
\renewcommand{\shorttitle}{The Power of 10: New Rules for the Digital World}

\hypersetup{
pdftitle={The Power of 10: New Rules for the Digital World},
pdfsubject={cs},
pdfauthor={Spiekermann-Hoff et al.},
pdfkeywords={Ethos, Digitalization, AI, Sustainability},
}

\begin{document}
\maketitle

\begin{abstract}
 As artificial intelligence rapidly advances, society is increasingly captivated by promises of superhuman machines and seamless digital futures. Yet these visions often obscure mounting social, ethical, and psychological concerns tied to pervasive digital technologies -- from surveillance to mental health crises. This article argues that a guiding ethos is urgently needed to navigate these transformations. Inspired by the lasting influence of the biblical Ten Commandments, a European interdisciplinary group has proposed ``Ten Rules for the Digital World''---a novel ethical framework to help individuals and societies make prudent, human-centered decisions in the age of ``supercharged'' technology.
 \end{abstract}


\section{Introduction}
It has become commonplace for tech companies to announce a breakthrough in Artificial Intelligence (AI) on an almost weekly basis. Many of these announcements claim to bring us closer to what has been termed ``Artificial General Intelligence (AGI)''---a level of AI that is supposed to move beyond the current sleight-of-hand of the various AI-powered chatbots and bring us machines that allegedly have ``superhuman abilities.'' (e.g., \cite{Kokotajlo2025}). In keeping with this rhetoric, human abilities are often at the heart of the names that tech companies choose for their products and services: machines are ``intelligent'' (instead of learning probabilistic classifiers), can ``think'' (instead of process), show ``emotion'' (instead of parsing data patterns statistically correlated with human affect), and are ``smart'' (instead of using sensors to provide context sensitivity or being simply connected to the Internet). It is no wonder that many engineers, managers, politicians, and private individuals engage in the envisioning and heralding of our future as a set-in-stone narrative of digital transformation that continues along the same lines of development we have pursued in the past three decades: technology will make our lives safer, more comfortable, more efficient, and happier.

At the same time, the relentless spread of information technology (IT) throughout society has not come without a cost: ever more citizen surveillance and data protection issues are surfacing from what \cite{Zuboff2019} calls a ``shadow economy.'' Powerful monopolistic platforms foster a relentless gig economy that not only undermines human potentials and adequate working conditions but also creates mental health issues and depression. In recent years, social and democratic skills are deteriorating as both adults and teenagers have become increasingly addicted to social media. In fact, heightened digital technology use has been linked to an increase in attention-deficit symptoms, impaired emotional and social intelligence, technology addiction, social isolation, impaired brain development, and disrupted sleep. The list of charges relating to the digital transformation seems endless, and just as advocates in Silicon Valley increasingly dream of ``enhancing'' humans with brain-computer interfaces and speculate about -superintelligence, a growing number of citizens and organizations are in truth suffering from computer errors.

In light of these developments, it seems that a general ethos would be needed regarding the question, how everyday engineering and system-use decisions could be taken more prudently. By an ethos we mean a short, comprehensive, easily memorisable, broadly interpretable, perhaps globally applicable everyday guide of behavior addressing not only a specific collective, such as engineers \citep{ACM2018}, a specific region (such as the EU), a specific technology (such as AI) or a specific problem space (such as the Digital Service Act) or industry, but individuals in all their roles at the individual and communitarian level. Such an ethos -- similar to the biblical Ten Commandments -- would help us to build and use technology more prudently everywhere and at any time and to become attentive to problem areas that expose us to unexpected reproaches and developments of the kind listed above. Is such an ethos at all feasible?

\section{The Design of a New Ethos}
In the fall of 2024, a group of renowned academics and professionals based in Europe joined forces and developed such an ethos for today’s digital society, calling it ``The 10 Rules for the Digital World''.\footnote{We want to thank all experts involved in the Future Foundation initiative and the Göttweig discussions. Beyond the authors themselves, this includes Oskar Aszmann, Christopher Coenen, Gerd Gigerenzer, Paul Nemitz, Walter Peissl, Matthias Pfeffer, Surjo Soekadar, Ludger Schwienhorst-Schönberger, Sigrid Stagl, Thomas Stieglitz, and Yvonne Hofstetter} All group members have spent most of their professional careers exploring ways in which different aspects of human, social and environmental issues can be better recognized and reflected in current technological and economic innovations. Their backgrounds cover the entire ``science fiction cycle'' of today’s AI technologies, including first-, second- and third-generation AI systems, as well as robotic and neural human enhancement technologies. In addition to computer science, their expertise spans a wide range of disciplines relevant in the context of accelerated digital transformation, such as medicine, technology assessment studies, law, military technology, neuroscience and bionics, psychotherapy, psychology, sociology, political science, socioeconomics, philosophy and theology.

Over the course of eight months, the group developed an ethos consisting of ten rules through a seven-stage process. This began with a two-day workshop in which IT and AI-related challenges were collected from twelve different expert perspectives. This problem-space collection triggered the insight that an ethos similar to the biblical Ten Commandments might be needed to tackle the enormous social and democratic challenges of the digital world. The group hence sought inspiration from this framework that -- arisen from the practical necessities of peaceful coexistence in the Middle East -- has been understood from the outset (and in later Jewish and Christian Theology) as the embodiment of a universal minimum ethical consensus \citep{Schockenhoff1996}. The Ten Commandments were in fact adapted in the Quran and are regarded by Muslims as part of a common ethical heritage of humanity \citep{Gunther2007}. Similar to non-Abrahamic traditions, such as Confucianism and the philosophical tradition of virtue ethics, they aim at character formation and not at the determination of voluntary acts by univocal rules (as in modern approaches to ethics like utilitarianism and deontology) \citep{Macintyre1985}. 

Against this background, three group members, consulting an expert in Old Testament studies, prepared a tabled framework describing the extended meaning of the Ten Commandments which was used as preparational guidance by the group in another two-day workshop to draft its first set of rules. A subsequent consolidation process following the workshop discussed how a meaningful correspondence could be exploited between the identified issues, the envisaged rules and the Ten Commandments, without connecting the 10 Rules for the Digital World to specific religious traditions. Further simplifications were made until the ten rules presented below were finalized. There were no objections against more religiously oriented members presenting the 10 Rules as closely linked to the Ten Commandments of the biblical tradition and other members relying exclusively on the values of democracy, fundamental rights as laid down in the UN Convention on Human Rights and the European Charter of Fundamental Rights, or principles of social justice.  

The 10 rules proposed for safeguarding our digital future are summarized in Table \ref{tab:ten} (a more extended version with a preamble is available at: \url{https://www.thefuturefoundation.eu}). Each entry reflects a key technological issue seen through the lens of one of the biblical Ten Commandments and then plays out two exemplary interpretations: the perspective of an ordinary end-user and that of a professional. By using the word ``interpretation,'' we signal that there are myriad ways to interpret a rule in a person’s individual context.

\begin{table}[ht]
  \caption{10 Rules for the Digital World}
  \resizebox{\textwidth}{!}{%
  \label{tab:ten}
  \begin{tabular}{ p{4cm} p{7cm} p{4cm} p{4cm}}
    \toprule
    Rule&Issues to debate and expand&Illustrative conclusions for end-users&Illustrative conclusions for professionals\\
    \midrule
    \bf 1. Do not elevate digital technology to an end in itself. 
    & The ideologization of technology for its own sake, including the lack of respect for the hierarchy of being, where the digital should serve the analogue and not vice versa. This also includes the tendency to fall in love with one’s own creations and promote the idea of ``superintelligence''. 
    & Avoid expecting technological solutions for every problem.
    & Refrain from striving to digitize everything.\\
    
    \bf 2. Do not wrongfully attribute humanity to machines. 
    & The use of imprecise, human-like naming of machine capabilities (anthropomorphism) and the construction or depiction of machines in ways that confuse technical capabilities with human qualities.
    & Refrain from building personal relationships with AI bots.
    & Avoid simulating a first-person perspective in AI conversations.\\
    
   \bf  3. Create space for downtime and analogue encounters.
   & The configuration of systems for permanent and non-discriminatory connectivity and reachability, along with the ubiquity of computing in all areas. This raises questions about the prioritization of efficiency over higher values such as leisure and well-being.
   & Establish regular routines for digital detox to create time and space away from screens.
   & Design systems that include spaces for leisure time and non-connectivity.\\
   
    \bf 4. Honor social and democratic capabilities. 
    & The replacement of meaningful social and family relationships with thousands of weak ties (e.g., ``friends'' on social media), as well as the lack of respectful and democratic discourse across ideological or social ``bubbles''.
    & Practice polite and healthy online behavior.
    & Promote and support polite and healthy communities both online and offline. \\
    
   \bf  5. Do not destroy nature for technological progress.
   & The recognition that IT (and AI specifically) is highly resource- and CO$_2$-intensive, prompting the need to revisit or balance the push for ubiquitous digital transformation in light of finite earthly resources and global ecological health.
   & Treat and utilize AI as a valuable and limited resource.
   & Develop IT solutions that prioritize minimal energy consumption.\\
   
  \bf   6. Do not treat people as mere data objects.
  & The potential for confusion between digital twins and their human counterparts, highlighting the importance of staying true and loyal to the person behind the user. This also includes concerns about sharing data on intimate analogue activities and the need to prioritize data minimization and quality.
  & Look beyond what the Internet tells you about people; seek deeper understanding.
  & Prioritize the individuality of people over computational scores.\\
  
  \bf   7. Do not deprive yourself and others of their true human potentials.
  & The recognition that human attention and creativity are among the most precious resources, requiring protection by design. This includes the careful orchestration of user accessibility to avoid default, disruptive, real-time, and ubiquitous access to people.
  & Leverage digital tools to support, not replace, your own creativity and knowledge.
  & Ensure that humans make the first creative proposal, allowing AI to improve it, not the other way around.\\
  
  \bf   8. Do not deny the limits of technology.
  & The need to remain realistic about the limitations of technological capabilities, including their error-proneness, security vulnerabilities, and added complexity. This also involves ensuring the protection of user rights in the face of machine errors.
  & Question machine-generated answers and decisions with confidence.
  & Acknowledge and address machine errors sincerely, and pursue error indications with vigilance.\\
  
  \bf   9. Do not undermine the freedom of others by technical means.
  & The risk of abusing technology to restrict the freedom of others, including the dangers of technology paternalism and mutual surveillance in private life.
  & Strike a careful balance between monitoring children or partners and respecting their right to freedom.
  & Avoid creating configurations (e.g., robots) that are physically restrictive and cannot be overridden.\\
  
  \bf   10. Prevent the concentration of power and ensure participation.
  & The monopolization or undue privatization of technological power or knowledge, as well as the excessive and non-democratically legitimized use of shared resources (e.g., water, energy, earth, space).
  & Support trustworthy local providers by paying for private computing resources.
  & Advocate for and support open-source, interoperable, modular, and controllable IT systems. \\
  \bottomrule
\end{tabular}}
\end{table}

\section{Making a Difference}
In recent years, the debate surrounding ethical AI, increased environmental sustainability awareness, and growing concerns about the democratic and social misuse of social networks and AI have all brought the subject areas covered by the proposed rules to the forefront of debate. There is now a general awareness of the threats that society faces from an unregulated digital world, and many regulators have started to react and address some of the issues covered by the rules (e.g., the EU AI Act \citep{AIAct2024}). So how are our rules different from or complementary to what has been done by others?

First, the 10 Rules encourage action in areas that are often overlooked, especially the first three. While some people seriously discuss the idea of computers becoming superintelligent, fewer openly criticize the tendency to replace living, dynamic systems with rigid, bureaucratic ones. Similarly, the push for more precise and honest language in science and tech marketing is rarely taken seriously, even though many criticize the issue informally. Efforts to rethink the always-on design of our systems are starting to gain attention in schools and workplaces, but this hasn’t yet been applied more broadly. These first three rules touch on deeper questions about what it means to be human and how technology should serve us, making it crucial to have open discussions and form clear opinions on these topics for shaping our future wisely. The use of the Ten Commandments as a structuring tool furthermore allowed recognizing the denial of technical limitations and the destruction of nature as additional issues.

Second, previous efforts have focused more on the organizational, technical or legal level. In academic disciplines like computer ethics, Value-Sensitive Design, Value-based Engineering, responsible innovation, participatory design, etc. many approaches have been developed as to how to design technology in a more responsible way. Many government agencies have established so-called ELSA/ELSI programs (ethical, legal and social aspects/implications), e.g., in the context of genomics research. Over eighty institutions have published principle lists for AI design \citep{Jobin2019}, including the AI4People Institute’s ``Good AI Society'' framework. However, these important efforts have not provided much direct guidance at the individual and community level, which is what our rules do, helping anyone in any role understand what is right and wrong. This is specifically illustrated in the two columns of table 1 with exemplary interpretations for individuals in their private roles as well as in their roles as designers of technology. For example, standardization bodies like those who originally determined the real-time response and always-on, stand-by mode of devices probably never imagined that their decision created a humanity that now lacks time and space for analog encounters. Parents using technology to monitor their children might not see how this surveillance could affect their child’s perception of freedom. And individuals using AI tools to create art or write text might not think about the energy costs involved or how relying on these tools could impact their own creativity in the long run.

Third, laws, regulations, industry standards, and organizational rules are limited by the borders of the countries or organizations to which they belong. Meanwhile, technology operates globally, often beyond the reach of these rules. A shared global ethos could act as a positive guiding force, helping us think more clearly about how we use and design technology in our daily lives.

Fourth, the 10 Rules that we propose here are short and easy to memorize. They are based on the ancient ``art of memory'' \citep{Yates1996}, something they have in common with the biblical Ten Commandments. While some scholars have recently started to argue for an ethical approach to computing with a specifically Christian focus for HCI design \citep{HinikerWobbrock2022}, those involved in creating these rules had mixed feelings about any particular religious framework. However, the authors of this paper still believe that the Ten Commandments have provided the group with a powerful inspiration for the global digital ethos proposed here.

Finally, and perhaps most importantly, the 10 Rules are open to interpretation. They do not dictate exactly what one should or should not do with a specific technology, in a specific context, a specific industry or specific region, but instead offer general guidance and a sense of direction to help people make better choices.

\section{The Need for Debate}
The rules we have proposed were created through a careful, collaborative process by recognized experts in technology research, who together bring over 320 years of combined professional experience to the table. However, given the enormous challenge of creating a global ethos, this is, of course, just a modest proposal.

While we believe that we have identified key issues for our digital future, there are many additional questions that could be investigated by future research and discourse. For example:  Do our rules highlight and cover the most important issues that need attention? Should the rules go into more depth and detail? And can the list be applied universally, across different cultures around the world or will different regions need their own interpretations? These are big questions that deserve thoughtful discussion and debate.

\bibliographystyle{unsrtnat}

\begin{thebibliography}{10}
\providecommand{\natexlab}[1]{#1}
\providecommand{\url}[1]{\texttt{#1}}
\expandafter\ifx\csname urlstyle\endcsname\relax
  \providecommand{\doi}[1]{doi: #1}\else
  \providecommand{\doi}{doi: \begingroup \urlstyle{rm}\Url}\fi

\bibitem[Kokotajlo et~al.(2025)Kokotajlo, Alexander, Larsen, Lifland, and Dean]{Kokotajlo2025}
Daniel Kokotajlo, Scott Alexander, Thomas Larsen, Eli Lifland, and Romeo Dean.
\newblock {AI 2027}.
\newblock \url{https://ai-2027.com/}, 2025.
\newblock AI Futures’ Project, Last Modified April 3rd, 2025, Accessed August 13th, 2025.

\bibitem[Zuboff(2019)]{Zuboff2019}
Shoshana Zuboff.
\newblock \emph{The Age of Surveillance Capitalism: The Fight for a Human Future at the New Frontier of Power}.
\newblock Public Affairs, New York, NY, 2019.

\bibitem[{ACM}(2018)]{ACM2018}
{ACM}.
\newblock {ACM Code of Ethics and Professional Conduct}.
\newblock \url{https://www.acm.org/code-of-ethics}, 2018.
\newblock Accessed August 29, 2025.

\bibitem[Schockenhoff(1996)]{Schockenhoff1996}
Eberhard Schockenhoff.
\newblock \emph{{Naturrecht und Menschenwürde. Universale Ethik in einer geschichtlichen Welt}}.
\newblock Matthias-Grünewald-Verlag, Mainz, 1996.
\newblock 261ff.

\bibitem[Günther(2007)]{Gunther2007}
Sebastian Günther.
\newblock {People of the Scripture! Come to a Word Common to You and Us (Q. 3:64): The Ten Commandments and the Qur’an}.
\newblock \emph{Journal of Qur'anic Studies}, 9\penalty0 (2):\penalty0 28--58, 2007.

\bibitem[Macintyre(1985)]{Macintyre1985}
Alasdair~C. Macintyre.
\newblock \emph{After virtue: {A} study in moral theory}.
\newblock Duckworth, London, 1985.

\bibitem[{European Union}(2024)]{AIAct2024}
{European Union}.
\newblock {Regulation (EU) 2024/1689 of the European Parliament and of the Council of 13 June 2024 laying down harmonised rules on artificial intelligence and amending Regulations (EC) No 300/2008, (EU) No 167/2013, (EU) No 168/2013, (EU) 2018/858, (EU) 2018/1139 and (EU) 2019/2144 and Directives 2014/90/EU, (EU) 2016/797 and (EU) 2020/1828}.
\newblock \emph{Official Journal of the European Union}, 2024\penalty0 (1689):\penalty0 144, 2024.
\newblock URL \url{https://eur-lex.europa.eu/legal-content/EN/TXT/PDF/?uri=OJ:L_202401689}.

\bibitem[Jobin et~al.(2019)Jobin, Ienca, and Vayena]{Jobin2019}
Anna Jobin, Marcello Ienca, and Effy Vayena.
\newblock The global landscape for {AI} ethics guidelines.
\newblock \emph{Nature Machine Intelligence}, 1:\penalty0 389--399, 2019.
\newblock \doi{10.1038/s42256-019-0088-2}.

\bibitem[Yates(1996)]{Yates1996}
Frances~A. Yates.
\newblock \emph{The Art of Memory}.
\newblock University of Chicago Press, London \& Chicago, 1996.

\bibitem[Hiniker and Wobbrock(2022)]{HinikerWobbrock2022}
Alexis Hiniker and Jacob~O. Wobbrock.
\newblock Reclaiming attention: {C}hristianity and {HCI}.
\newblock \emph{Interactions}, 29\penalty0 (4):\penalty0 40--44, 2022.

\end{thebibliography}

\end{document}